\documentclass{ws-ijmpa}
\usepackage[super,compress]{cite}
\usepackage{graphicx}
\usepackage{hyperref}
\begin{document}
\markboth{Felix Kahlhoefer}{Review of LHC Dark Matter Searches}

%
\catchline{}{}{}{}{}
%

\title{Review of LHC Dark Matter Searches}

\author{Felix Kahlhoefer}

\address{DESY, Notkestra\ss e 85, D-22607 Hamburg, Germany\\
felix.kahlhoefer@desy.de}

\maketitle


\begin{abstract}

This review discusses both experimental and theoretical aspects of searches for dark matter at the LHC. An overview of the various experimental search channels is given, followed by a summary of the different theoretical approaches for predicting dark matter signals. A special emphasis is placed on the interplay between LHC dark matter searches and other kinds of dark matter experiments, as well as among different types of LHC searches.

\end{abstract}

\ccode{PACS numbers: 95.35.+d, 12.60.-i, 13.85.Rm\\Preprint number: DESY-17-024}

\tableofcontents

\section{Introduction}

At first sight it may seem preposterous to search for dark matter (DM) at hadron colliders. After all, we have so far only observed the gravitational interactions of DM and it therefore remains unknown whether DM particles interacts sufficiently strongly with ordinary matter to enable us to produce them in collisions of Standard Model (SM) particles. Even if such a production is possible in principle, the unknown DM mass means we do not know whether the center-of-mass energy available in present experiments is sufficient for the production to be kinematically allowed. Finally, should the production of DM particles succeed against these odds, the resulting experimental signature would be most unspectacular: The DM particles would be invisible to any detector in the vicinity of the collision point and would hence reveal their presence only via an apparent imbalance in the total transverse momentum (so-called missing transverse momentum or, more colloquially, missing energy).

Nevertheless, searches for DM particles at the Large Hadron Collider (LHC) are a thriving research field, which have attracted significant interest from both the experimental and the theoretical community in recent years. The main reason is the experimental feasibility of these searches. The SM backgrounds and their differential distributions are now understood to sufficient accuracy that even small distortions in the missing energy spectrum (in particular in its high-energy tail) may be observable and can be used to constrain DM models that predict these distortions.

This leads to the question of how to construct DM models for predicting missing energy signals at the LHC. Clearly, any such model must contain a new neutral stable particle consistent with the properties inferred from the behavior of DM on astrophysical scales (regarding e.g.\ structure formation). This requirement alone is however too general to provide useful guidance for experiments and to yield instructive results. Instead, the fundamental requirement should be that these models provide a mechanism for how DM was produced in the early Universe, so that the predictions of the model can be compared to the one well-measured property of DM, namely its cosmological relic density $\Omega_\text{DM} \, h^2 \approx 0.12$~\cite{Ade:2015xua}.

One of the most successful paradigms for DM production in the early Universe (in the sense of its predictivity and its ability to reproduce the observed relic density) is the idea of thermal freeze-out~\cite{Lee:1977ua}. This idea is based on the assumption that DM particles interact sufficiently strongly with ordinary matter that they enter into thermal equilibrium with the bath of SM states at high temperatures. The relic abundance is then essentially set by the temperature at which the DM annihilation rate drops below the expansion rate of the Universe and therefore the DM interactions become insufficient to maintain thermal equilibrium -- the DM particles freeze out.

\begin{figure}[t]
\centering
\includegraphics[width=0.7\textwidth]{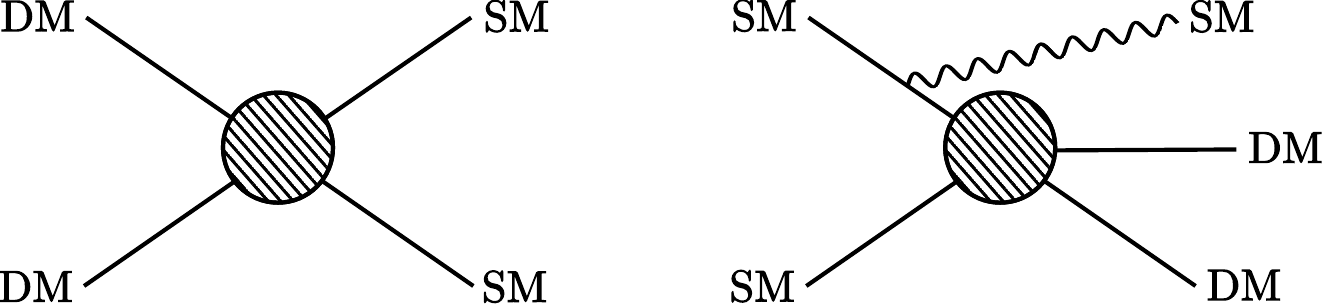}
\caption{Illustration of the connection between thermal freeze-out and DM searches at the LHC. The observed DM relic abundance can be readily reproduced if the cross section for DM annihilation (left) is sizeable. It should then be possible to invert this annihilation process by colliding SM particles. To obtain an observable DM signal at the LHC, it is necessary to produce DM in association with SM states (right).}
\label{fig:DMproduction}
\end{figure}

In principle, the freeze-out mechanism can work for a wide range of masses and couplings.\footnote{In practice there is an upper bound on the DM mass of about $m_\text{DM} \lesssim 100\:\text{TeV}$ from the requirement of perturbative unitarity~\cite{Griest:1989wd} and in many models one finds lower bounds of $m_\text{DM} \gtrsim 10\:\text{GeV}$ from the requirement that DM thermal freeze-out does not spoil the successful predictions of recombination~\cite{Ade:2015xua}. } The crucial point is however that in any case interactions between the DM particle and SM states have to be sizable. Specifically, the velocity-averaged DM annihilation cross section should roughly be given by
\begin{equation}
\langle \sigma_{\text{DM DM} \to \text{SM SM}} \, v \rangle \approx 3 \times 10^{-26} \mathrm{cm^3 \, s^{-1}} \; . 
\end{equation}
The generic expectation would then be that the inverse process also has a sizable cross section and that therefore particle colliders can be used to invert the annihilation processes that happened frequently in the early Universe, see figure~\ref{fig:DMproduction}. This strong link to the idea of thermal freeze-out justifies the excitement for DM searches at the LHC.

Another key reason to search for DM at the LHC is that new stable particles at the electroweak scale are a central ingredient of models that attempt to address the gauge hierarchy problem, such as supersymmetry~\cite{Jungman:1995df}. Intriguingly, if the DM mass is comparable to the electroweak scale (i.e.\ of order of a few hundred GeV) and the coupling strength is comparable to that of the weak interactions, the required DM annihilation cross section can be obtained rather naturally.\footnote{Such DM candidates are often referred to as weakly-interacting massive particles (WIMPs) and the successful prediction of their relic abundance is sometimes called the \emph{WIMP miracle}. } In such a set-up the LHC is expected to produce DM particles in abundance via the decays of heavier (colored) states with large production cross sections, leading to characteristic missing energy signatures. Many searches for these well-motivated extensions of the SM are therefore ultimately also searches for DM.

Nevertheless, in models like supersymmetry, DM particles are typically only produced together with a significant number of additional SM particles from the decay chain, implying that there is no direct connection between the annihilation and the production process.\footnote{On the other hand, the presence of these additional states means that it may be much easier to infer the properties of the DM particle, such as its mass, from kinematic distributions.} Furthermore, most constraints on these models are in fact unrelated to the actual properties of the DM particle. In the context of these searches, the DM particle is therefore often a mere tool rather than the actual object of interest.

This review will therefore adopt a more narrow definition of LHC DM searches and focus on \emph{searches for the direct pair-production of DM making use of SM particles emitted from an initial, intermediate or final state}.\footnote{For the same reason we will also not discuss the (very interesting) signatures arising if DM is produced in the decay of long-lived particles.} The review is structured as follows. First, in section~\ref{sec:signatures} the relevant experimental signatures are discussed and some recent results are presented. Section~\ref{sec:models} then provides an overview of different approaches for predicting the expected DM signals. Finally, section~\ref{sec:complementarity} focuses on the important issue of how DM searches at the LHC can be connected to other LHC searches, as well as to non-collider DM searches, in order to obtain complementary information.

\section{Experimental searches}
\label{sec:signatures}

In this section we review the experimental strategies relevant for DM searches at the LHC and provide references to the most recent experimental results. The main focus of this discussion is on so-called \emph{mono-$X$ searches}, which are concerned with the production of a single SM particle in association with missing transverse momentum, but some more complicated DM signatures are also mentioned. To avoid repetition, it is worth pointing out that at the time of writing none of the DM searches discussed below have observed a significant excess over expected backgrounds.

\subsection*{Mono-jet}

If it is at all possible to produce DM particles in proton-proton collisions, it should also be possible to produce them in association with one or more QCD jets from initial state radiation. Searches for events in which a jet with high transverse momentum $p_T$ is produced in association with large missing transverse momentum $E_{T,\rm miss}$ have therefore become emblematic for LHC DM searches. The popular name \emph{mono-jet search} is however rather misleading, because the probability to produce just one highly energetic jet is in fact rather low~\cite{Haisch:2013ata}. Mono-jet searches therefore typically only impose a strict veto on events containing leptons, but do include events with several high $p_T$ jets.

In fact, with increasing center-of-mass energy these searches have become more and more inclusive. For example, the most recent ATLAS analysis allows up to four jets with $p_T > 30\:\text{GeV}$ and pseudorapidity $|\eta| < 2.8$, while the leading (i.e.\ most energetic) jet is required to have $p_T > 250\:\text{GeV}$ and $|\eta| < 2.4$~\cite{Aaboud:2016tnv}. CMS does not constrain the total number of jets at all and only requires that the leading jet satisfy $p_T > 100\:\text{GeV}$ and $|\eta| < 2.5$~\cite{CMS:2016pod}.

Clearly it becomes very challenging to model distributions of missing transverse momentum in events with such a large number of jets, making it necessary to use data-driven methods based on control regions for the background estimation. For example, the pertinent background from $p p \to Z(\to \nu \bar{\nu}) + \text{jets}$ can be inferred from analogous events in which the $Z$ boson decays leptonically. Similarly, backgrounds arising from $p p \to W (\to \ell \nu) + \text{jets}$ and an unobserved charged lepton can be inferred from events in which the lepton is correctly identified. 

Detector effects, in particular jet mismeasurement, can also lead to events that appear to have unbalanced transverse momentum. However, these multi-jet backgrounds can be very efficiently suppressed by requiring that the missing transverse momentum vector does not point into the (azimuthal) direction of any of the leading jets. In combination, these methods allow to describe the background distributions rather well. Nevertheless, the sensitivity of mono-jet searches is often limited by systematic uncertainties and improving it with increasing luminosity is rather challenging. To make progress it will be essential to better understand the impact of electroweak corrections on the $W + \text{jets}$ to $Z + \text{jets}$ ratio~\cite{Denner:2012ts}.

In some models the DM particles are produced preferentially in association with one or more bottom quarks~\cite{Lin:2013sca,Agrawal:2014una}. Searches for such models are conceptually similar to mono-jet searches, except that they require at least one hard jet to pass $b$-tagging requirements~\cite{Aad:2014vea,ATLAS-CONF-2016-086,CMS:2016uxr}.

\subsection*{Mono-$V$}

In a similar manner as in mono-jet events, DM may also be produced together with a vector boson $V = \gamma, W, Z$, which is radiated off a quark in the initial state. While the corresponding production cross section is significantly smaller than for QCD radiation, the process is much cleaner and can therefore be searched for with higher sensitivity. Moreover, if DM particles couple directly to a pair of gauge bosons, mono-$V$ processes may in fact be the dominant way in which DM is produced at the LHC~\cite{Carpenter:2012rg}.

Mono-photon searches are among the conceptually simplest searches for DM, requiring typically only the presence of a high $p_T$ photon and no isolated leptons~\cite{Aaboud:2016uro,CMS:2016fnh}. Although both detector effects (for example electron or jet misidentification) and beam-induced events can potentially fake mono-photon events, background levels are typically very low and the experimental sensitivity is therefore limited only by statistics.

Leptonically decaying $Z$ bosons also yield a very clean signal~\cite{ATLAS:2016bza,CMS:2016hmx,Sirunyan:2017onm}. By requiring that the transverse momentum of the di-lepton system is opposite in azimuthal direction and similar in magnitude to the missing transverse momentum vector and that the di-lepton invariant mass is close to the $Z$ boson mass, backgrounds can be very much suppressed and only the irreducible backgrounds from di-boson production remain relevant. 

If a $W$ boson produced in association with DM decays leptonically, the neutrino adds to the missing transverse momentum and one obtains a so-called mono-lepton event~\cite{Khachatryan:2014tva,ATLAS:2014wra}. Since the experimental signature is essentially identical to the one from the leptonic decay of an off-shell $W$ boson, background suppression is challenging and requires an accurate estimate of the transverse mass distribution.

It is also possible to search for the production of a $W$ or $Z$ boson in association with missing transverse momentum in the hadronic final state. Searches for hadronically decaying $W$ or $Z$ bosons are similar to mono-jet searches but use a larger distance parameter for the leading jet and employ additional criteria such as requiring the mass of the fat jet to be consistent with a $W$ or $Z$ boson~\cite{Aaboud:2016qgg,CMS:2016pod}.

We note that mono-$W$ searches have received significant interest in the context of so-called \emph{isospin-violating DM}, i.e.\ DM particles with different couplings to up and down quarks~\cite{Giuliani:2005my}. This interest results from the observation that a typical mono-$W$ signal arises from the interference of two different diagrams, in which DM couples to up quarks and down quarks, respectively. Mono-$W$ searches should therefore have a unique sensitivity to the relative phase of the two couplings. However, it was subsequently pointed out that it is very difficult to consistently study this set-up without violating gauge invariance~\cite{Bell:2015sza}. We will return to this issue in section~\ref{sec:consistency}.

\subsection*{Mono-Higgs}

Searches for a SM-like Higgs boson in association with missing transverse momentum were first proposed in phenomenological studies~\cite{Petrov:2013nia,Carpenter:2013xra,Berlin:2014cfa} and have since triggered a number of experimental searches both in the $\gamma \gamma$~\cite{Aad:2015yga,CMS:2016xok,ATLAS-CONF-2016-011} and in the $b\bar{b}$~\cite{Aad:2015dva,Aaboud:2016obm,CMS:2016mjh} final state. In the former case, backgrounds are very small and hence only a relatively loose cut on $E_{T,\rm miss}$ is necessary. For example, the most recent CMS search~\cite{CMS:2016xok} requires $E_{T,\rm miss} > 105\:\text{GeV}$. Moreover, these searches can make use of the excellent resolution in the invariant mass of the photon pair, $m_{\gamma\gamma}$, to suppress non-resonant backgrounds from SM processes with mismeasured $E_{T,\rm miss}$. As a result, mono-Higgs searches in the di-photon channel are currently only limited by statistics.

In the $b\bar{b}$ final state, on the other hand, background rejection is of crucial importance. Fortunately, these searches can draw from a number of techniques developed to identify Higgs bosons with high $p_T$. A particularly interesting situation occurs in the case that the SM Higgs boson is produced with sufficiently high transverse momentum that the two $b$-jets from its decay merge into a single fat jet. Very roughly, this is the case if $2 m_h / p_T < R$, where $R$ is the distance parameter of the fat jet clustering algorithm. For example, the ATLAS collaboration searches for events with $E_{T,\rm miss} > 500\:\text{GeV}$ in which there is a single fat jet with $R = 1.0$ and $p_T > 250\:\text{GeV}$ that contains two $b$-tagged sub-jets with $R = 0.2$~\cite{Aaboud:2016obm}.  This procedure achieves a Higgs tagging efficiency of up to $40\%$~\cite{ATLASperformance}. Again, the dominant backgrounds ($t\bar{t}$ and $Z,W+\text{jets}$) are non-resonant, so that the invariant mass of the fat jet can be used to discriminate signal from background. These searches are still very much limited by statistics.

\subsection*{DM + top quarks}

If DM particles couple dominantly to heavy quark flavors, a promising way to discover DM at the LHC is to search for a top-quark pair in association with missing transverse momentum~\cite{Lin:2013sca,Arina:2016cqj}. Both ATLAS~\cite{Aad:2014vea} and CMS~\cite{CMS:2016mxc} have performed such searches in the channel where at least one of the $W$ bosons from the decay $t \to bW$ decays into hadrons. Top quarks produced in association with DM are typically not highly boosted, so their decay products can be fully resolved. A number of cuts on the kinematics of the individual jets as well as $b$-tagging techniques can then be used to efficiently suppress backgrounds. In addition, CMS has also performed a search in the fully leptonic channel~\cite{CMS:2016jxd}, requiring at least two jets, out of which at least one is $b$-tagged, as well as exactly two leptons.

While searches in the semi-leptonic channel provide the strongest constraints at the moment~\cite{Haisch:2015ioa}, searches in the fully-leptonic channel are nevertheless interesting for a number of reasons. First of all, these searches are currently only limited by statistics and therefore promise significant gains in sensitivity with luminosity~\cite{Haisch:2016gry}. Furthermore, the angular distribution of the two leptons can be used not only to distinguish signal from background, but also to determine the CP nature of the DM interactions~\cite{Haisch:2013fla,Haisch:2016gry}.

Finally, it is also conceivable that DM is produced in association with a single top quark~\cite{Andrea:2011ws}, and corresponding searches have been carried out by both ATLAS~\cite{Aad:2014wza} (in the semi-leptonic channel) and by CMS~\cite{Khachatryan:2014uma,CMS:2016flr} (in the fully hadronic channel). It is worth noting, however, that (due to the absence of top quarks in the initial state) this signature can only arise either from sizable flavor-changing transitions, or from a bottom quark in the initial state analogous to single-top production~\cite{Pinna:2017tay}.

\subsection*{Invisible Higgs decays}

If the DM mass is less than half of the mass of the SM Higgs boson, it may be possible to produce pairs of DM particles in Higgs decays. Such invisible Higgs decays can be searched for in a number of different ways. First of all, indirect constraints can be obtained by combining the visible decay modes in order to construct an upper bound on all unobserved decay channels. This approach, however, requires an assumption on the Higgs production cross section, which is typically taken to be given by the SM prediction.

Alternatively, one can directly search for invisible Higgs decays by triggering on the particles that signal the production of a Higgs boson. The two most relevant production modes in this context are vector boson fusion (VBF) and production in association with a massive vector boson (VH). However, the gluon-fusion mode can also be interesting if the Higgs boson is produced with an additional jet from initial state radiation.

Searches for the associated production of a Higgs boson together with a massive vector boson followed by an invisible decay of the Higgs boson are conceptually very similar to the mono-$V$ signatures discussed above. They can be searched for both in the leptonic decays of a $Z$ boson or in the hadronic decays of a $W$ or $Z$ boson~\cite{Khachatryan:2016whc,Aad:2015uga,ATLAS:2016bza}. The jet-associated gluon-fusion production mode, on the other hand, essentially yields a mono-jet signature as discussed in detail above~\cite{Khachatryan:2016whc}.

A truly novel signature is obtained in the VBF case~\cite{Rainwater:1997dg}. In this channel the production of a Higgs boson is signaled by the presence of two jets with large separation in pseudo-rapidity and large invariant mass~\cite{Aad:2015txa,Khachatryan:2016whc}. This distinctive topology can be exploited to discriminate hypothetical invisible Higgs decays from the large SM backgrounds. Indeed, searches for invisible Higgs decays in the VBF production mode typically yield the strongest bounds on the Higgs invisible branching ratio.

Combinations of the various direct searches for invisible Higgs decays have been performed by both ATLAS~\cite{Aad:2015pla} and CMS~\cite{Khachatryan:2016whc}, the latter including first results from data taken at 13 TeV. The resulting upper bounds at 95\% confidence level on the invisible branching ratio are $\text{BR}_\text{inv} < 0.25$ and $\text{BR}_\text{inv} < 0.24$, respectively. These bounds are comparable to the ones obtained indirectly from the visible decay modes~\cite{Aad:2015pla,Khachatryan:2016vau}.

\section{Predicting dark matter signals}
\label{sec:models}

Having discussed the various ways in which one can search for the production of DM at the LHC, we now turn to the essential question: How do we know what a DM signal at the LHC will look like? Which of the search channels is the most promising? And how do we calculate the details of the expected distributions? It should be clear that (for the time being) there is no single correct answer to these questions. Our ignorance of the particle physics nature of DM means that a variety of different approaches need to be considered. Each approach needs to find a compromise between two conflicting requirements: generality and plausibility (or equivalently minimality and realism)~\cite{DeSimone:2016fbz}.

The requirement of generality (or minimality) means that we want to make as few assumptions as possible on the presence of new particles in the dark sector. This approach is desirable both from the practical perspective (because only a small number of new parameters are introduced) and from the philosophical perspective (along the lines of Occam's razor). The requirement of plausibility (or realism), on the other hand, compels us to give preference to models that are in agreement with well-established principles of particle physics, for example the absence of large $\mathrm{CP}$ or flavor violation. This requirement may also include aesthetic or practical arguments, for example that the model under consideration should have a perturbative ultra-violet (UV) completion that is consistent with the structure of the SM gauge group before electroweak symmetry breaking.

Another way to characterize these two conflicting approaches is to describe them as the bottom-up approach on the one side, in which we try to add the minimum amount of additional structure to the SM, and the top-down approach on the other side, in which we start from well-motivated UV completions in order to gain intuition for the construction of DM models. We will begin with the bottom-up approach below and then extend this approach further and further until we connect to models obtained from the top-down approach.

\subsection{Effective theories}

The most minimal assumption possible is that the DM particle is the only new state beyond the SM that is kinematically accessible at the LHC. In this case, the interactions between DM and SM states can be described by an effective field theory (EFT) containing operators of mass dimension larger than four.\footnote{There are a small number of DM models that pursue an even more minimal approach by coupling the DM particle to the SM via renormalizable interactions, such as sterile neutrinos~\cite{Abazajian:2012ys}, hidden photons~\cite{Redondo:2008ec}, scalar singlets~\cite{Silveira:1985rk,McDonald:1993ex,Burgess:2000yq} or $SU(2)_L$ multiplets with a stable neutral component~\cite{Cirelli:2005uq}. With the exception of the scalar singlets, which we will discuss below in the context of Higgs portal models, these models do however not predict any observable signals at the LHC.} This approach was first suggested under the name \emph{Maverick Dark Matter}~\cite{Beltran:2010ww} and was subsequently popularized as the \emph{EFT approach} by a number of detailed studies~\cite{Goodman:2010yf,Bai:2010hh,Goodman:2010ku,Fox:2011fx,Fox:2011pm,Rajaraman:2011wf}. 

An extensive classification of the lowest-dimension effective operators describing the interactions between fermionic or scalar DM and quarks or gluons has been performed in refs.~\citen{Goodman:2010yf,Goodman:2010ku}.\footnote{The relevance of higher-dimension operators is discussed in ref.~\citen{Bruggisser:2016ixa}. } Each operator is characterized by only two parameters: the effective suppression scale $\Lambda$ and the DM mass $m_\text{DM}$. For example, a frequently studied operator is the so-called axial-vector operator:
\begin{equation}
\label{eq:axialvector}
 \mathcal{O} = \frac{1}{\Lambda^2} (\bar{q} \gamma^\mu \gamma^5 q)(\bar{\chi} \gamma_\mu \gamma^5 \chi) \; ,
\end{equation}
where $\chi$ denotes a spin-1/2 DM particle, which can be either a Dirac fermion (in which case the operator is usually labeled D8~\cite{Goodman:2010ku}) or a Majorana fermion (labeled M6~\cite{Goodman:2010yf}). This operator has been the subject of a number of LHC studies~\cite{Aad:2015zva,Khachatryan:2014rra}.

A crucial property of the EFT approach is that the shape of all kinematic distributions is independent of the suppression scale $\Lambda$. For a dimension-6 operator proportional to $\Lambda^{-2}$, for example, all cross sections are simply proportional to $\Lambda^{-4}$. It is therefore technically very easy to present experimental results in terms of lower bounds on $\Lambda$ as a function of $m_\text{DM}$. For DM masses smaller than the typical cut on missing transverse momentum, kinematic distributions and hence the resulting bounds become independent of $m_\text{DM}$, implying that LHC searches can be sensitive to arbitrarily small DM masses.

For hadron colliders, the strongest constraints on the suppression scale $\Lambda$ are obtained for effective operators involving quarks and gluons. Another interesting possibility however are contact interactions between DM particles and SM gauge bosons~\cite{Carpenter:2012rg,Cotta:2012nj,Crivellin:2015wva} or Higgs bosons~\cite{Carpenter:2013xra}. In such a set-up, any gauge boson or Higgs boson produced at the LHC can radiate off a pair of DM particles, potentially leading to mono-$V$ or mono-Higgs signals.

Effective interactions between DM particles and Higgs bosons have also been studied in the context of so-called Higgs portal models~\cite{Baek:2011aa,Djouadi:2011aa,Djouadi:2012zc,LopezHonorez:2012kv}. Indeed, one of the simplest ways to couple fermionic DM to the SM is via the dimension-5 operator
\begin{equation}
 \mathcal{O} \sim \frac{1}{\Lambda} H^\dagger H \, \bar{\chi}\chi\;,
 \label{eq:higgsportal}
\end{equation}
where $H$ denotes the SM Higgs doublet. After electroweak symmetry breaking, this operator gives rise to an $h\bar{\chi}\chi$ vertex, where $h$ denotes the physical Higgs boson. For $m_\text{DM} < m_h /2$ this interaction leads to invisible Higgs decays, which are strongly constrained by experimental data. For $m_\text{DM} > m_h/2$, on the other hand, the DM production cross section at the LHC is strongly suppressed~\cite{Craig:2014lda}. We note that, if DM is a scalar singlet, the corresponding Higgs portal operator is in fact renormalizable, leading to the arguably simplest model for DM production at the LHC.

\subsubsection*{EFT validity}

The original appeal of the EFT approach was based on the idea that bounds on effective operators are model-independent, in the sense that it is not necessary at any point of the analysis to specify the details of the underlying UV completion. However, this hope has been challenged by two related observations. First, it has become clear that there are many interesting models describing the production of DM at the LHC which are not correctly captured by the EFT approach~\cite{Bai:2010hh,Fox:2011pm,Fox:2012ee}. In other words, these models predict kinematic distributions that differ significantly from the ones obtained from contact interactions. And second, it was shown that~-- at least for certain values of the suppression scale $\Lambda$~-- the effective operator approach makes unphysical predictions so that it becomes impossible to find a plausible UV completion~\cite{Shoemaker:2011vi,Fox:2012ee,Endo:2014mja}.

Both of these observations are connected to the way in which effective operators are obtained from a more fundamental theory. For example, the axial-vector operator from eq.~(\ref{eq:axialvector}) can be obtained from a theory containing a heavy spin-1 particle $V^\mu$ with axial couplings to DM and quarks:
\begin{equation}
 \mathcal{L} \supset \frac{m_V^2}{2} V^\mu V_\mu + V^\mu g_q \bar{q} \gamma_\mu \gamma^5 q + V^\mu g_\text{DM} \bar{\chi} \gamma_\mu \gamma^5 \chi \; .
\end{equation}
In the context of DM searches at the LHC, such a new particle is often referred to as the \emph{mediator} of the interactions between quarks and DM. If the mediator is exchanged in the $s$-channel of a $2\rightarrow2$ process with center-of-mass energy $\sqrt{s}$, the resulting matrix element will contain a propagator of the form
\begin{equation}
 \mathcal{M} \propto \frac{g_q \, g_\text{DM}}{m_V^2 - s} \; .
\end{equation}
In the limit $m_V^2 \gg s$ this propagator becomes $g_q \, g_\text{DM} / m_V^2$ and one obtains the axial-vector operator from above with
\begin{equation}
\label{eq:lambda}
 \frac{1}{\Lambda^2} = \frac{g_q \, g_\text{DM}}{m_V^2} \; .
\end{equation}
If, on the other hand, the mass of the mediator is comparable to or smaller than the momentum transfer in the process, the contact interaction does not provide an accurate description of the kinematics, because terms that are of higher order in $s / m_V^2$ cannot be neglected. In other words, if we are interested in a theory where DM interacts with quarks via the exchange of a mediator with mass at or below the TeV scale, the EFT approach will not correctly capture this model at LHC energies.\footnote{We emphasize that it becomes more complicated to apply this argument to effective operators for which the physical interpretation of the suppression scale is less clear, such as the one introduced in eq.~(\ref{eq:higgsportal}).}

As long as $m_V$ is sufficiently large, the EFT approach is fully justified at LHC energies. However, while the validity of the EFT approach depends on the value of $m_V$, the sensitivity of experimental searches depends only on the suppression scale $\Lambda$. It turns out that the LHC has no sensitivity to effective operators involving DM with a suppression scale larger than a few TeV. In fact, to obtain sizable cross sections $\Lambda$ typically has to be below the TeV scale. At first sight, this does not necessarily pose a problem: If the couplings $g_q$ and $g_\text{DM}$ are very large, it follows from eq.~(\ref{eq:lambda}) that $m_V$ can be much larger than $\Lambda$ and hence the EFT approach may still be valid.  

The problem we then face is that the matrix element obtained from the effective operator scales proportional to $s / \Lambda^2$ with increasing center-of-mass energy (or equivalently that the cross section grows proportional to $s / \Lambda^4$). At some point the EFT therefore violates the requirement of perturbative unitarity, which essentially demands that the matrix element (or, more precisely, all of its partial waves) is smaller than unity~\cite{Shoemaker:2011vi,Fox:2012ee,Endo:2014mja}. Typically, this happens for $\sqrt{s} \sim (2\text{--}3) \Lambda$.

The conclusion is that for large momentum transfer the EFT makes unphysical predictions. In other words, the energy transfer for at least some fraction of events at the LHC is sufficient to resolve the underlying micro-physics. For a complete description of all processes at the LHC the effective operator must then be replaced by a more complete theory, in which unitarity is restored~\cite{Bell:2016obu}.

\subsubsection*{EFT truncation}

A possible way to restore the validity of the EFT approach for suppression scales $\Lambda$ comparable to LHC energies is to ensure that the EFT is only applied to processes with sufficiently small momentum transfer that the predictions of the effective operator can be trusted. This process, referred to as \emph{EFT truncation}, allows to obtain a conservative but model-independent bound.

The most straightforward way to perform such a truncation is to assume that the EFT approach becomes invalid at energies $E \gtrsim g_\ast \Lambda $~\cite{Racco:2015dxa,Bruggisser:2016nzw}. In the example discussed above, this scale would be given by $m_V$, i.e.\ one can identify $g_\ast = \sqrt{g_q \, g_\text{DM}}$. For a specific choice of $\Lambda$ and $g_\ast$ one then disregards all events that have a momentum transfer larger than $g_\ast \Lambda$ and determines whether the remaining number of events is sufficient to exclude the assumed value of $\Lambda$.\footnote{Since the DM particles in the final state cannot be detected, the actual momentum transfer in a given event is not observable, so this requirement can only be implemented in Monte Carlo generators.} Iterating this procedure, one can construct a self-consistent bound on $\Lambda$ for an assumed value of $g_\ast$~\cite{Aad:2015zva}. An alternative approach~\cite{Busoni:2013lha,Berlin:2014cfa,Busoni:2014sya} is to determine the fraction of events with $E > g_\ast \Lambda$ and to use this number to rescale bounds obtained without truncation. Note that, since the energy required to produce a pair of DM particles is at least $E = 2 \, m_\text{DM}$, it is never possible to constrain suppression scales with $g_\ast \Lambda < 2 \, m_\text{DM}$.

If the underlying model of DM is strongly interacting, i.e.\ if $g_\ast \gg 1$, the EFT approach together with an appropriate truncation procedure yields the most suitable description of DM production at the LHC~\cite{Bruggisser:2016nzw}, avoiding the problem that the EFT makes unphysical predictions. If, on the other hand, DM is weakly coupled, so that $g_\ast \lesssim 1$, kinematic distributions are typically so different from the EFT approach that it is not possible to obtain relevant constraints using effective operators. Clearly, a different description of the interactions of DM is necessary. This alternative approach is provided by the framework of DM simplified models.

\subsection{Simplified models}

So far we have discussed possible ways to describe the interactions of DM particles at the LHC under the assumption that they are the only particles kinematically accessible. Now we will go one step further and consider possible descriptions in which there is a second light particle, which is responsible for mediating the interactions of quarks and DM. To limit the number of possibilities, we now require all interactions to be renormalizable, i.e. we will only consider interactions of dimension four or less. This approach~\cite{Buchmueller:2013dya} has come to be known as \emph{DM simplified models}~\cite{Abdallah:2014hon}.

Clearly, including a light mediator in the model requires the introduction of a number of new parameters, on which kinematic distributions will depend in a non-trivial way. While this additional complexity is a challenge for experimental searches, there are a number of good reasons to consider such a set-up:
\begin{enumerate}
 \item There is no simple way to translate an exclusion limit obtained within the EFT approach to models with a light mediator. A naive conversion of a bound on $\Lambda$ into a bound on the mediator mass $m_\text{med}$ may both overestimate or underestimate the actual strength of the constraint by orders of magnitude~\cite{Buchmueller:2013dya}. Simplified models aim to fill this gap and provide bounds for models that cannot be mapped onto effective operators.
 \item Contact interactions between DM particles and SM states predict rather specific distributions, in particular very hard missing energy spectra. If experiments relied only on these predictions for developing analysis strategies, the searches would likely not have the optimum sensitivity for models predicting softer spectra. Considering mediators of different masses makes it possible to consider different kinematic distributions and optimize the experimental sensitivity for each case.
 \item From a theoretical point of view, it is quite natural to assume that the DM particle is comparable in mass to the particle responsible for its interactions. In particular it turns out to be very difficult to obtain the required DM relic abundance if the mediator is too heavy, challenging one of the primary motivations for LHC DM searches. In the presence of a light mediator, on the other hand, the relic abundance can be readily reproduced.~\cite{Busoni:2014gta}
\end{enumerate}
Note that, while the name suggests a top-down motivation, simplified models are really most strongly motivated from a bottom-up perspective. We will return to the question of how these models are connected to well-motivated UV completions in section~\ref{sec:consistency}.

Following a joint effort of the LHC experiments and the theoretical community, a number of particularly interesting simplified models have been identified~\cite{Abdallah:2015ter} and their LHC phenomenology has been investigated in the context of the ATLAS/CMS DM Forum (DMF)~\cite{Abercrombie:2015wmb}. These simplified models have subsequently been the subject of a number of theoretical studies~\cite{Buchmueller:2014yoa,Harris:2014hga,Buckley:2014fba,Xiang:2015lfa,Harris:2015kda,Brennan:2016xjh} and experimental analyses~\cite{Aaboud:2016tnv,CMS:2016pod}.

\subsubsection*{$s$-channel mediators}

We have already mentioned a first example for such a simplified model above, namely the axial-vector mediator
\begin{equation}
 \mathcal{L} \supset g_q \, V^\mu \sum_q \bar{q} \gamma_\mu \gamma^5 q + g_\text{DM} \, V^\mu \bar{\chi} \gamma_\mu \gamma^5 \chi \; .
\end{equation}
Analogously, one can consider the case of a vector mediator\footnote{Note that the vector current vanishes for Majorana fermions and for real scalars. Consequently, the interactions between DM and a vector mediator can only sensibly be defined if the DM particle is a Dirac fermion or a complex scalar.}
\begin{equation}
 \mathcal{L} \supset g_q \, V^\mu \sum_q \bar{q} \gamma_\mu q + g_\text{DM} \, V^\mu \bar{\chi} \gamma_\mu \chi \; .
\end{equation}
Note that, to be consistent with the hypothesis of minimal flavor violation, the mediators are assumed to couple to all quarks with equal strength~\cite{Abdallah:2015ter}. Because of the way how DM is produced in these simplified models, they are referred to as spin-1 $s$-channel models. In a similar way one can construct spin-0 $s$-channel models. The case of a scalar mediator $\phi$ is given by
\begin{equation}
 \mathcal{L} \supset g_q \, \phi \sum_q \frac{y_q}{\sqrt{2}} \bar{q} q + g_\text{DM} \, \phi \bar{\chi} \chi \; ,
\end{equation}
while the pseudoscalar mediator $a$ is described by
\begin{equation}
 \mathcal{L} \supset g_q \, a \sum_q \frac{y_q}{\sqrt{2}} \bar{q} \gamma^5 q + g_\text{DM} \, a \bar{\chi} \gamma^5 \chi \; .
\end{equation}
The fact that these interactions are taken to be proportional to the Yukawa couplings $y_q$ is again a result of the hypothesis of minimal flavor violation~\cite{Abdallah:2015ter}. In the context of spin-0 $s$-channel mediators, DM is therefore expected to couple most strongly to top quarks~\cite{Haisch:2013fla,Buckley:2015ctj,Haisch:2016gry,Arina:2016cqj,Pinna:2017tay}. As we will discuss below, $g_q$ should be thought of as a numerical factor (such as a mixing angle) rather than an independent coupling.

Each of the four simplified models described above is fully characterized by five parameters: the masses of the two particles $m_\text{DM}$ and $m_\text{med}$, the two couplings $g_q$ and $g_\text{DM}$ and the width of the mediator $\Gamma_\text{med}$. In fact, the shapes of the various kinematic distributions depend only on the two masses and the mediator width, whereas a variation of the couplings simply rescales the spectra. Varying the couplings while keeping the width fixed may however lead to unphysical results if the total width is assumed to be smaller than the sum of the partial widths for decays into DM and quarks. To avoid this problem, it is usually assumed that the mediator does not couple to any other light particles, so that its width can be calculated in terms of the remaining four parameters (often referred to as the \emph{minimal width assumption}):
\begin{equation}
 \Gamma = \Gamma_{\chi\bar{\chi}} + \sum_q \Gamma_{q\bar{q}} + \Gamma_{gg} \; ,
\end{equation}
where the final term is relevant only for spin-0 $s$-channel mediators.

Replacing the mediator width by the minimal width eliminates one parameter, but it has the disadvantage that now the shapes of the different spectra depend on all four remaining parameters in a non-trivial way. Nevertheless, in many cases the dependence on the couplings is trivial, whereas the two masses turn out to be decisive for the experimental sensitivity. It has therefore become common to study simplified models by considering specific (fixed) choices of couplings and investigate the experimental sensitivity in the parameter plane spanned by the two masses~\cite{Boveia:2016mrp}.

\subsubsection*{Presenting experimental results}

\begin{figure}[t]
\centering
\includegraphics[width=0.5\textwidth,clip,trim=0 20 0 20]{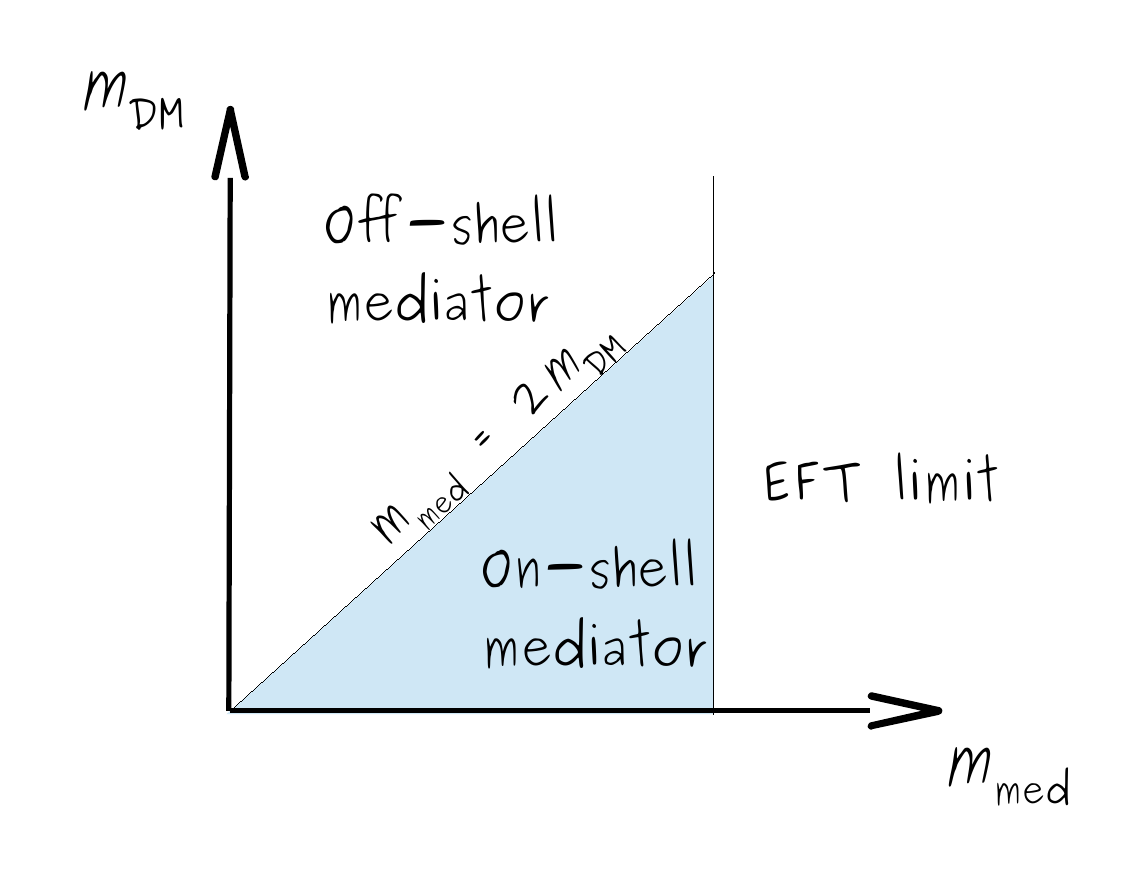}
\caption{Sketch of the mass-mass plane used to present experimental results for simplified DM models. See text for details.}
\label{fig:sketch}
\end{figure}

A sketch of such a study is shown in figure~\ref{fig:sketch}. Roughly, the mass-mass plane can be divided into three different regions. For very large mediator masses, one recovers the EFT limit discussed above. LHC searches typically have no sensitivity to this case unless the couplings are assumed to be close to the perturbativity bound. For smaller mediator masses, the phenomenology depends decisively on the ratio of the mediator mass and the DM mass. For $m_\text{med} > 2 m_\text{DM}$, the mediator can be produced on-shell and subsequently decays into a pair of DM particles. As a result, the DM production cross section receives a resonant enhancement. Conversely, for $m_\text{med} < 2 m_\text{DM}$ the production of DM can only proceed via an off-shell mediator and is correspondingly suppressed. To first approximation, LHC searches therefore aim to explore the on-shell region.

Presenting experimental results in the mass-mass plane is very convenient for comparing the sensitivity of different LHC searches. In addition, it is possible to indicate in this plane the parameter combinations for which the total DM annihilation cross section is equal to the thermal cross section, so that the observed DM relic abundance can be reproduced (see section~\ref{sec:complementarity}). The drawback of this approach is that it is far from obvious how to reinterpret a specific experimental result in terms of a model with couplings that differ from the assumed values (or with additional contributions to the mediator width).

One possible solution is to make use of the narrow-width approximation, which states that in the on-shell region the cross section for any mono-$X$ process factorizes into the production cross section of the mediator together with the state $X$ and the branching ratio of the mediator to decay into DM particles:
\begin{equation}
 \mathrm{d} \sigma(p p \to \chi \bar{\chi} + X) = \mathrm{d} \sigma(p p \to V + X) \cdot \text{BR}(V \to \chi \bar{\chi}) \; .
\end{equation}
In particular, this approximation, which is valid as long as $\Gamma_\text{med} \lesssim 0.3 \, m_\text{med}$, implies that the shape of kinematic distributions depends \emph{only} on $m_\text{med}$. It is then possible to infer a bound on any model containing an $s$-channel mediator as follows
\begin{enumerate}
 \item For the model of interest, calculate the branching ratio $\text{BR}_0 = \text{BR}(V \to \chi \bar{\chi})$.
 \item Identify the simplified model most similar to the model of interest, pick a parameter point within the on-shell region with the same mediator mass and read off the bound on the signal strength $\mu$ (i.e. the ratio of excluded cross section to predicted cross section).
 \item For the chosen parameter point in the simplified model, calculate the branching ratio $\text{BR}_1 = \text{BR}(V \to \chi \bar{\chi})$.
 \item The bound on the signal strength for the model of interest is then given by $\bar{\mu} \equiv \mu \frac{\text{BR}_1}{\text{BR}_0}$. For $\bar{\mu} < 1$, the model is excluded by the search under consideration.
\end{enumerate}
This procedure provides a good approximation to more detailed analyses as long as the parameter point under consideration is well within the on-shell region and the mediator width is sufficiently narrow~\cite{Jacques:2015zha,Brennan:2016xjh}.

\subsubsection*{$t$-channel mediators}

Another well-motivated class of simplified models focuses on the case that the mediator couples to one quark and one DM particle. Clearly, such an interaction is only possible if the mediator carries color charge. A frequently studied case is the one where the DM particle is a fermion and the mediator is a colored scalar:
\begin{equation}
 \mathcal{L} \supset g \sum_{i=1,2,3} \phi_i^\ast \bar{\chi} P_R u_i \; .
\end{equation}
Interactions with right-handed down-type quarks and left-handed quarks can be constructed in complete analogy.

In order to be consistent with minimal flavor violation, the mediator needs to carry a flavor index $i = 1,2,3$, implying that there are really three mediators of equal mass ($m_1 = m_2 = m_3$) and equal coupling strength $g_1 = g_2 = g_3 \equiv g$. It is possible, however, to break this universality and consider the case where the third-generation mediator has different mass and couplings from the first two generations. Most studies then focus on the two mediators coupling to the first two generations. These models are referred to as \emph{$t$-channel flavored mediators}~\cite{Bell:2012rg,Chang:2013oia,An:2013xka,Bai:2013iqa,DiFranzo:2013vra,Papucci:2014iwa,Garny:2014waa,Busoni:2014haa,Garny:2015wea}. Another interesting possibility, named \emph{flavored dark matter} is that the DM particle (rather than the mediator) carries the flavor index~\cite{Agrawal:2011ze,Agrawal:2014una}. Models in which the third-generation mediator dominates the phenomenology are often called \emph{top-flavored dark matter}~\cite{Kumar:2013hfa,Batell:2013zwa,Kilic:2015vka}.

The mediator in this set-up resembles very much the squarks in the minimal supersymmetric Standard Model and indeed the collider phenomenology is very similar. Compared to the case of an $s$-channel mediator, there are additional contributions to mono-jet signals, because the mediator can radiate off gluon jets or decay into a jet and a DM particle. Moreover, it is possible to pair-produce the mediator, leading to a distinctive signature with two jets and two DM particles in the final state: $p p \to \phi \phi^\ast \to \chi \bar{\chi} j j$~\cite{An:2013xka,Bai:2013iqa,DiFranzo:2013vra,Papucci:2014iwa}. Searches for mono-jets and di-jets in association with missing transverse momentum have comparable sensitivity, unless the mass spectrum is highly degenerate (i.e. $m_\text{med} - m_\text{DM} \ll m_\text{med}$), in which case mono-jet searches are more promising~\cite{Papucci:2014iwa}.

To conclude the discussion of simplified models, we point out that there are a number of further possibilities, for example those involving scalar DM particles or fermionic mediators. For a more complete overview (as well as comprehensive lists of analytical results), we refer to more focused reviews of DM simplified models~\cite{Abdallah:2014hon,DeSimone:2016fbz}. For a discussion of simplified models with spin-2 mediators, we refer to ref.~\citen{Kraml:2017atm}.

\subsection{Towards complete models}
\label{sec:consistency}

The central idea of the simplified model framework is that, while it abandons some of the model independence of the EFT approach, it still captures the phenomenology of a wide range of possible theories of DM. The aim is therefore that one can obtain relevant constraints for any such theory by mapping it onto the appropriate simplified model. It should be clear that in some cases the constraints obtained in this way may not be the dominant ones, meaning that the simplified model may not capture all of the relevant phenomenology. However, at the very least this approach should make it possible to conclusively rule out certain regions of parameter space based on the predictions of the corresponding simplified model.

While the technical details of such a mapping from complete theories to simplified models remain challenging (see above for a discussion of how this can be done in the narrow width approximation), this section will focus on a more fundamental question, namely whether the simplified models introduced above are the appropriate choice for this purpose. Indeed, it is perfectly conceivable that the models introduced above are too simplified, in the sense that we have neglected additional states or couplings that make an essential contribution to the phenomenology. Conversely, it is also possible that we have been too general. For example, it may turn out that there are hidden relations between the different parameters, or that certain regions of parameter space are disfavored for theoretical reasons.

In principle one could try to address these questions from a top-down approach by considering a large collection of UV-complete models of DM and performing the mapping onto simplified models. Clearly, if a specific simplified model is never found to give a relevant constraint, its inclusion in LHC studies would be questionable. However, it is possible to address the same issue also from the bottom-up perspective, by investigating whether the simplified models introduced above fulfill certain theoretical consistency requirements, such as gauge invariance and that perturbative unitarity is guaranteed in the relevant regions of parameter space~\cite{Kahlhoefer:2015bea,Bell:2015rdw}. As we will see below, these principles provide a useful guidance for the construction of more realistic simplified models.

\subsubsection*{Implications of perturbative unitarity}

One of the main motivation for the development of simplified models was to address the issue of unitarity violation ubiquitous in the EFT approach. Indeed, it has been shown that the simplified models introduced above satisfy the requirement of perturbative unitarity in mono-jet and mono-$Z$ searches up to very large energies~\cite{Englert:2016joy}, provided all couplings are sufficiently small. Nevertheless, some cases have been identified in which (seemingly renormalizable) simplified models are in fact not well-behaved up to arbitrarily high energies.

A particularly interesting case are mono-$W$ searches. For simplified models with an $s$-channel spin-1 mediator, these searches are found to violate perturbative unitarity at high energies unless DM couples with equal strength to left-handed up and down quarks~\cite{Bell:2015sza,Bell:2015rdw,Haisch:2016usn}. This unphysical behavior turns out to result from the emission of longitudinal $W$ bosons and can only be tamed if the simplified model is extended by an additional interaction between the mediator and $W$-bosons~\cite{Haisch:2016usn}. A similar solution is found in models with a $t$-channel colored scalar mediator, where the emission of $W$-bosons from the mediator itself restores unitarity at high energies~\cite{Bell:2015rdw}.

The simplified model with an axial-vector mediator turns out to be troublesome for yet another reason: In contrast to a spin-1 mediator with purely vectorial couplings, the longitudinal mode of the axial-vector mediator does not decouple. In fact, if the transverse mode couples to a fermion of mass $m_f$ with coupling strength $g_f$, the corresponding coupling of the longitudinal mode is proportional to $g_f \, m_f / m_\text{med}$, i.e.\ it is enhanced for heavy fermions. In other words, for $m_f \gg m_\text{med}$ the coupling of the mediator will become non-perturbative.\footnote{In practice, this is not an issue for mono-jet searches, where the mediator always couples to a light quark in the initial state~\cite{Englert:2016joy}. This situation changes, however, when considering searches for top quarks in association with missing transverse momentum, where all fermions involved in the process can be heavy~\cite{Chala:2015ama,Englert:2016joy,Kahlhoefer:2015bea}. } This consideration implies that the DM mass should satisfy the requirement~\cite{Kahlhoefer:2015bea}
\begin{equation}
 m_\text{DM} \leq \sqrt{\frac{\pi}{2}} \frac{m_\text{med}}{g_\text{DM}} \; ,
\end{equation}
giving an example for how theoretical considerations can constrain the parameter space of simplified models.

But even if all couplings are perturbative, the simplified model with an axial-vector mediator still violates unitarity at high energies (for example in the process $\chi \bar{\chi} \rightarrow V V$, which yields a matrix element proportional to $\sqrt{s}$ for large energies)~\cite{Kahlhoefer:2015bea}. This issue is reminiscent of the well-known problems with unitarity that the SM would face in the absence of a Higgs boson. Clearly, the inconsistencies of the axial-vector simplified model arise from the fact that we have not specified a mechanism to generate the mediator mass. The simplest way to address this issue is to introduce an additional Higgs boson that is a singlet under the SM gauge group~\cite{Kahlhoefer:2015bea,Bell:2016uhg}. This so-called \emph{dark Higgs boson} then acquires a vacuum expectation value that generates the mediator mass.\footnote{For a spin-1 mediator with purely vectorial couplings it is possible to generate the mediator mass via a Stueckelberg mechanism~\cite{Stueckelberg:1900zz} without the need to introduce additional degrees of freedom.}

The observation that simplified models with an axial-vector mediator are necessarily incomplete has led to an increasing interest in simplified models with more than one mediator~\cite{Ghorbani:2015baa,Choudhury:2015lha,Duerr:2016tmh}. Although it is necessary to introduce at least one additional parameter (the mass of the second mediator), these models are highly attractive due to their rich phenomenology. For example, if the spin-1 mediator decays visibly, one may obtain a mono-$Z'$ signature~\cite{Autran:2015mfa,Bai:2015nfa}, while visible decays of the spin-0 mediator may lead to a mono-dark-Higgs signal~\cite{Duerr:2017uap}.

Another consistency requirement, which is relevant for any UV completion of spin-1 simplified models that involves an extension of the SM gauge group, is the absence of gauge anomalies. One possible way to implement this requirement is to choose the couplings of the mediator to SM fermions in such a way that no anomalies arise~\cite{Liu:2011dh,Ekstedt:2016wyi,Jacques:2016dqz}. Doing so necessarily implies that the mediator couples to leptons, leading to tight constraints from searches for di-lepton resonances (see below). Insisting on a mediator that couples only to quarks, on the other hand, requires new states that cancel the anomalies~\cite{Duerr:2013dza,Perez:2014qfa,Ismail:2016tod}. Nevertheless, in many cases there is no color anomaly~\cite{Kahlhoefer:2015bea} and therefore no new colored states are required in order to achieve anomaly freedom. The additional states are therefore expected not to affect the phenomenology of the model significantly.\footnote{The detailed particle content of the model does however become important when considering loop-induced processes such as DM annihilation into gamma-rays~\cite{Duerr:2015vna}. }

\subsubsection*{Implication of gauge invariance}

The example of mono-$W$ searches discussed above illustrates the problems that may arise if a simplified model does not respect the full gauge symmetry of the SM before electroweak symmetry breaking. It may nevertheless be of interest from a phenomenological perspective to study simplified models with interactions that can only arise after electroweak symmetry breaking. An instructive example are simplified models with an $s$-channel scalar or pseudoscalar mediator. In these models the mediator is a SM singlet which couples to $\bar{q} q = \bar{q}_L q_R + \bar{q}_R q_L$ and $\bar{q} \gamma^5 q = \bar{q}_L q_R - \bar{q}_R q_L$, respectively. Clearly, neither of these interactions is invariant under the SM gauge group before electroweak symmetry breaking, so they must vanish in the limit that the electroweak vacuum expectation value tends to zero and electroweak symmetry is restored~\cite{Bell:2015sza}.

A possible way to obtain these interactions is to assume that the spin-0 mediator does not actually couple directly to SM quarks, but that it only obtains these couplings after electroweak symmetry breaking from mixing with the SM Higgs~\cite{Bell:2016ekl}.\footnote{Note that for a pseudoscalar mediator, such a mixing would violate $\mathrm{CP}$ symmetry~\cite{Ghorbani:2014qpa,Ghorbani:2016edw,Baek:2017vzd}. } In this case, the new mediator couples to fermions in exactly the same way as the SM Higgs, i.e.\ with coupling strength proportional to the fermion masses. This construction therefore leads to the same coupling structure that was motivated above by invoking minimal flavor violation.

At first sight, such a \emph{simplified model with mixing} gives rise to a very interesting phenomenology. The fact that the mediator couples most strongly to heavy quarks means that gluon fusion via top-quark loops will give the dominant contribution to its production cross section, leading to promising mono-jet signals~\cite{Haisch:2012kf,Buckley:2014fba,Harris:2014hga,Haisch:2015ioa}. Moreover, the mediator may be produced in association with heavy quarks~\cite{Lin:2013sca,Artoni:2013zba}, which may potentially allow to distinguish scalar from pseudoscalar mediators~\cite{Buckley:2015ctj,Haisch:2016gry}. Finally, the mediator would also obtain couplings to SM gauge bosons, so that even mono-$V$ signals can be expected.

In practice, however, all of these promising signatures are forced to be small by the measurements of the branching ratios of the SM-like Higgs boson. Even if the DM particle is sufficiently heavy that invisible Higgs decays are kinematically forbidden, bounds on the Higgs signal strength require the mixing angle between the SM Higgs and the new mediator to be rather small. In other words, gauge invariance implies that the most relevant constraints on the simplified model with mixing will likely come from Higgs physics rather than from LHC DM searches. 

A possible way to evade this conclusion is to extend the Higgs sector with a second Higgs doublet. The mediator that couples to DM can then obtain its couplings to SM quarks from mixing with the second Higgs doublet.\footnote{An alternative approach is to consider an inert second Higgs doublet, which does not have any direct couplings to SM fermions. The lightest component of this inert doublet then is a candidate for scalar DM, which obtains couplings to SM states via the Higgs portal~\cite{LopezHonorez:2006gr}. } This way, rather than modifying the branching ratios or the signal strength of the SM Higgs boson, the mixing will modify the properties of the second Higgs doublet (and may thereby even explain why no heavier Higgs boson has yet been observed). Since the second Higgs doublet contains both a scalar and a pseudoscalar degree of freedom, it is easily possible within this approach to generate couplings between a pseudoscalar singlet mediator and SM quarks without violating $\mathrm{CP}$~\cite{Ipek:2014gua,No:2015xqa}.

Simplified models based on two Higgs doublet models have received significant interest recently~\cite{Ipek:2014gua,No:2015xqa,Goncalves:2016iyg,Bauer:2017ota}. This interest stems from the observation that the additional heavy Higgs bosons may lead to novel signatures not captured by the simplified model with just a single mediator. In particular, the heavy Higgs bosons may decay into the singlet mediator and a SM $Z$ or Higgs boson. If the singlet mediator then decays into DM particles, one may obtain large mono-$Z$ or mono-Higgs signals. Furthermore, it may be possible to produce the mediator in the VBF mode and perform searches analogous to the ones relevant for constraining invisible Higgs decays (see section~\ref{sec:signatures})~\cite{Brooke:2016vlw}.

\bigskip

In the discussion above we have moved from the largely model-independent EFT approach towards simplified models that predict a rich phenomenology and finally to various extensions of these models that address potential theoretical inconsistencies. The resulting variety of approaches provides a continuous spectrum from the most minimal constructions to highly complex models. Indeed, some of the extended simplified models discussed above could equally well have been obtained from a top-down approach. For example, simplified models containing a spin-1 mediator and a dark Higgs are very similar to models of $U(1)'$ extensions of the SM gauge group~\cite{Duerr:2013dza,Duerr:2014wra}, whereas simplified models containing two Higgs doublets and a SM singlet spin-0 mediator resemble the Higgs sector of the next-to-minimal supersymmetric Standard Model. Clearly, it is desirable to pursue all of these different approaches in parallel, so that the balance between generality and plausibility can be adjusted for each situation as appropriate.

\subsection{Numerical tools}

The effective theories and simplified models discussed above provide a framework for predicting the various signatures of DM production at the LHC discussed in section~\ref{sec:signatures} and can therefore be used to devise the corresponding experimental searches and to present the resulting constraints. To achieve this goal it is necessary to implement these models into numerical codes for Monte Carlo event generation capable of making accurate predictions for collider observables.

Generating DM signals at next-to-leading order (NLO) is essential in particular for signatures where the DM particle is produced in association with QCD radiation. In the context of the EFT approach these QCD NLO corrections have been calculated~\cite{Fox:2012ru} and implemented into the Monte Carlo generator MCFM~\cite{Campbell:2010ff}. This approach has subsequently been extended to $s$-channel simplified models\cite{Haisch:2013ata} and implemented into the POWHEG BOX~\cite{Alioli:2010xd}. All DMF $s$-channel simplified models have also been implemented into FeynRules~\cite{Alloul:2013bka} and MadGraph5\_aMC@NLO~\cite{Alwall:2014hca} at the NLO level under the name DMsimp~\cite{Mattelaer:2015haa,Backovic:2015soa,Neubert:2015fka}. Very recently, model files for some extended simplified models have also become available~\cite{Bauer:2016gys,Bauer:2017ota}.

Finally, we note that there are dedicated numerical tools for calculating DM observables beyond the LHC, such as micrOMEGAs~\cite{Belanger:2014vza}, MadDM~\cite{Backovic:2015cra} and DarkSUSY~\cite{Gondolo:2004sc}. An ongoing effort to combine all of this information into global fits of DM models is performed by the GAMBIT collaboration~\cite{Cornell:2016gho}. This way it will be possible to fully explore the complementarity of different DM searches, which we will discuss in the next section.

\section{A question of complementarity}
\label{sec:complementarity}

\begin{figure}[t]
\centering
\includegraphics[width=0.9\textwidth,clip,trim = 0 0 150 0]{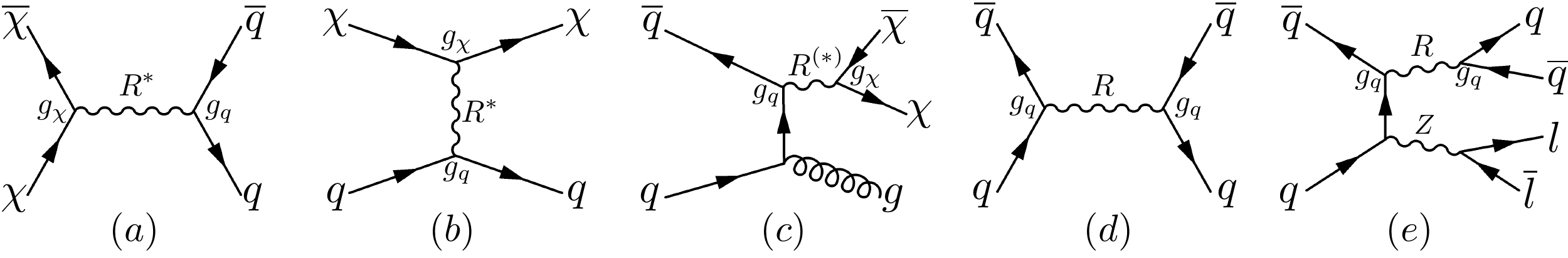}
\caption{Illustration of the different processes involving a DM particle $\chi$ and an $s$-channel mediator~$R$. From left to right the diagrams correspond to DM annihilation, DM-nucleon scattering, DM production together with a jet from initial state radiation and the decay of the mediator into a di-jet resonance.}
\label{fig:diagrams}
\end{figure}

The different approaches for predicting DM signals at the LHC presented in the previous section all have in common that they can be used not only to obtain the rates for DM production in the collisions of SM particles, but also to calculate the cross sections for other types of processes involving DM particles. Of particular interest in this context are the DM annihilation cross section, which is relevant for calculating the DM relic abundance and for indirect detection experiments, and the cross section for DM-nucleon scattering, which determines the event rates in direct detection experiments. Figure~\ref{fig:diagrams} illustrates this observation for the case of a simplified model with an $s$-channel mediator.

If the same model can be used to obtain predictions for several types of experiments, we can use it to compare the sensitivity of the different experimental strategies and to map bounds from one kind of search onto the parameter space relevant for another approach. Such a comparison can be important for a number of reasons. As long as there is no conclusive signal, we may be interested in understanding which experimental strategy is the most promising and where we would expect to first see a DM signal. Furthermore, we may hope to learn how large the allowed parameter space is, telling us what sensitivity we need to aim for and when to stop pursuing a specific search.

The possibility to compare different search strategies will become even more important once a conclusive mono-$X$ signal has been observed at the LHC. Crucially, the LHC alone cannot establish the stability of invisible particles it produces. To infer the DM nature of such particles necessarily involves the connection to non-collider experiments and to cosmological observations. It is therefore essential for any \emph{DM search} at the LHC (as opposed to a mere search for invisible particles) to be constructed in a way that facilitates such a connection.

A detailed discussion of how the comparison between different experimental strategies is made for the different approaches discussed above is beyond the scope of this review (some of the possible caveats are discussed in refs.~\citen{Profumo:2013hqa,Bauer:2016pug}). Instead, we will discuss the general approach and some of the inherent complications, and present some of the key results.

\subsection{Relic density and indirect detection}

As discussed in the introduction, one of the central motivations for DM searches at the LHC stems from the idea of thermal freeze-out, which motivates sizable interactions between DM particles and SM states. It is therefore a pertinent question whether this link holds in practice, i.e.\ whether the LHC achieves sufficient sensitivity to test the freeze-out paradigm. 

Within a specific model, it is straightforward to calculate the DM annihilation cross section and compare it to the thermal cross section $\langle \sigma v \rangle = 3 \cdot 10^{-26} \: \mathrm{cm^3 s^{-1}}$ \mbox{$\approx 2.5 \cdot 10^{-9} \: \text{GeV}^{-2}$}. If, for example, DM interacts with quarks via a dimension-6 effective operator with suppression scale $\Lambda$, we expect from dimensional arguments an annihilation cross section of the order of $m_\text{DM}^2 / \Lambda^4$ (assuming $m_\text{DM} \gg m_q$). If the DM particle is light (i.e.\ of order of a few tens of GeV or less) this estimate implies that $\Lambda$ must be well below the TeV scale (and therefore within the reach of colliders) in order to achieve a sufficiently large annihilation cross section. Collider constraints therefore typically severely constrain thermal production for low-mass DM~\cite{MarchRussell:2012hi,Cheung:2012gi,Haisch:2013uaa,Busoni:2014gta}.

There are a number of caveats to this line of reasoning. First of all, LHC DM searches primarily probe the interactions of DM with quarks and gluons. If DM interacts primarily with color-neutral SM states such as leptons, it will be very difficult for the LHC to constrain the coupling strengths implied by thermal freeze-out. Furthermore, there are many ways in which the standard thermal production mechanism can be altered. For example, there could be late-time entropy injection (reducing the DM abundance) or a particle-antiparticle asymmetry in the dark sector (increasing the DM abundance)~\cite{MarchRussell:2012hi}. Finally, there could be additional particles in the dark sector, so that the DM particle can experience co-annihilation~\cite{Baker:2015qna,Buschmann:2016hkc,Khoze:2017ixx}. Nevertheless, it remains interesting to understand whether standard thermal production via the couplings between DM and quarks can yield the observed relic abundance or whether LHC bounds imply the presence of some additional ingredient (in the form of additional couplings or modifications of standard cosmology) to match observations. 

The same processes that set the DM relic abundance can be searched for in indirect detection experiments. If DM annihilates into quarks, the ensuing hadronization produces a continuum of $\gamma$ rays as well as a characteristic flux of anti-protons, which can potentially be observed with satellites such as Fermi-LAT~\cite{Ackermann:2015zua} or AMS-02~\cite{Aguilar:2016kjl}. The bounds obtained from these searches can then be mapped onto the parameter space relevant for LHC searches~\cite{Goodman:2010qn,Zheng:2010js,Cheung:2011nt,Duerr:2015wfa,Esmaili:2016enf}.
Conversely, one can use the simplified model framework to compare LHC constraints with astrophysical excesses to infer whether a consistent DM interpretation is possible~\cite{Buchmueller:2015eea,Balazs:2015boa}.

It is important to keep in mind that in contrast to LHC searches, indirect detection experiments rely on the presence of DM in astrophysical objects (such as the Galactic Center or Milky Way dwarf spheroidals) and therefore suffer from astrophysical uncertainties related to the DM distribution. We will return to this issue in the context of direct detection experiments.

\subsection{Direct detection}

Direct detection experiments aim to detect the scattering of DM particles off nuclei in low-background underground detectors. Since DM particles in the Galactic halo move with velocities of order $v \sim 10^{-3}$, the momentum transfer in a DM scattering event is typically $\mu v \lesssim 100\:\text{MeV}$ with $\mu$ being the DM-nucleus reduced mass~\cite{Peter:2013aha}. DM particles therefore interact not with individual quarks but coherently with the entire nucleus. As a result there are two main types of scattering: spin-independent scattering and spin-dependent scattering. In the former case the DM particle couples to the mass $A$ (or charge $Z$) of the entire nucleus and scattering rates therefore receive a coherent enhancement proportional to $A^2$ (or $Z^2$). In the latter case, the DM particle couples dominantly to unpaired nucleons within the nucleus, so that there is no large enhancement factor~\cite{Boveia:2016mrp}.

DM direct detection experiments thus place very strong bounds on any model of DM that predicts spin-independent interactions~\cite{Tan:2016zwf,Akerib:2016vxi}, in particular simplified models with vector or scalar mediators as well as the corresponding effective operators~\cite{Buchmueller:2014yoa}. For spin-dependent interactions, on the other hand, constraints are comparable in strength to the ones obtained from the LHC for a wide range of DM masses. While direct detection is more sensitive to DM in the TeV range (where the LHC runs out of energy), the LHC has a distinct advantage for low-mass DM. The reason is that for DM masses below a few GeV, the momentum transfer in direct detection experiments becomes too small to be detectable for existing experimental strategies. 

An alternative way to probe spin-dependent scattering is to consider the capture of DM particles in the sun, followed by DM annihilation~\cite{Silk:1985ax}. Typically, these two processes are in equilibrium, so that the annihilation rate is determined by the capture rate and hence directly comparable to direct detection experiments. If the DM annihilation process leads to the production of neutrinos (for example as a result of hadronization), the capture rate can be constrained by neutrino telescopes such as IceCube~\cite{Aartsen:2016exj} or Super-Kamiokande~\cite{Choi:2015ara}. Indeed, these constraints are comparable to the ones obtained by the most sensitive direct probes of spin-dependent scattering~\cite{Amole:2015pla,Amole:2016pye} and provide complementary constraints on DM simplified models~\cite{Heisig:2015ira,Jacques:2016dqz}.

On closer inspection, it turns out that the simple division into spin-independent and spin-dependent interactions does not capture all possibilities. A general approach in terms of a non-relativistic effective theory of DM scattering identifies twelve additional operators that predict smaller but potentially relevant scattering rates~\cite{Fitzpatrick:2012ix,Anand:2013yka,Bishara:2016hek}. It is possible to construct a mapping from the various simplified models to this set of operators and hence calculate event rates in direct detection experiments for all of the models discussed in section~\ref{sec:models}~\cite{Dent:2015zpa}. It turns out that some of these models predict a strong suppression of event rates in the non-relativistic limit and are therefore completely unconstrained by direct detection experiments. This is for example the case for simplified models with a pseudoscalar mediator, which have received significant interest for precisely this reason~\cite{Boehm:2014hva,Dolan:2014ska}. 

A further subtlety in the mapping from collider constraints to direct detection experiments results from the large separation of scales between the energies relevant for LHC collisions and the ones describing the scattering of DM particles from the Galactic halo. Indeed, it has been shown that renormalization group evolution can lead to important effects, in particular for models where the tree-level calculation predicts a strong suppression of the event rates~\cite{Frandsen:2012db,Haisch:2013uaa,Crivellin:2014qxa,Crivellin:2014gpa,DEramo:2014nmf,DEramo:2016gos,Bishara:2016hek}. 

Unfortunately, the comparison between LHC results and direct detection experiments is complicated by a number of theoretical uncertainties. First of all, calculating event rates in direct detection experiments requires knowledge of the relevant nuclear form factors, which can carry uncertainties of $10\%$ or more~\cite{Cline:2013gha}. Even larger uncertainties result from our lack of knowledge of the local DM density and the corresponding velocity distribution~\cite{Fairbairn:2012zs}. There is consequently a great need for techniques to compare LHC and direct detection experiments without the need to make strong assumptions on the astrophysical parameters (so-called halo-independent methods)~\cite{Ferrer:2015bta,Blennow:2015gta}. Finally, it is important to keep in mind that direct detection experiments rely on the assumption that all of DM consists of only one type of particles. If there are instead several different sub-components, direct detection experiments will be at a disadvantage compared to the LHC~\cite{Chala:2015ama}.

\subsection{Searches for mediators}

In addition to the complementarity between the various DM searches discussed above, there is an important interplay between all these searches and processes not involving any DM particles at all. For example, if DM can be produced at the LHC via an $s$-channel mediator, this mediator may always decay back into quarks or gluons (see panel (d) in figure~\ref{fig:diagrams})~\cite{An:2012va,Frandsen:2012rk,An:2012ue}. Consequently searches for di-jet resonances place strong bounds on models for DM production at the LHC~\cite{Fairbairn:2014aqa,Chala:2015ama,Fairbairn:2016iuf,ATLAS:2016bvn,ATLAS:2016lvi,Sirunyan:2016iap}. A similar argument can be made in the EFT approach. If for example DM particles interact with quarks via an effective operator of the form $\frac{1}{\Lambda^2} \bar{q}\gamma^\mu q \, \bar{\chi} \gamma_\mu \chi$, one would in general also expect the presence of an analogous operator involving only quarks: $\frac{1}{\Lambda'^2} \bar{q} \gamma^\mu q \, \bar{q} \gamma_\mu q$ with $\Lambda' \approx \Lambda$~\cite{Dreiner:2013vla}.

In the context of spin-1 $s$-channel simplified models, the strong constraints from searches for di-jet resonances imply that sizable mono-$X$ signals can only be expected if the mediator couples much more strongly to DM than to quarks~\cite{Chala:2015ama,Fairbairn:2016iuf}. In this case, there is an intriguing interplay between di-jet searches and mono-$X$ searches: In the on-shell region, where $m_\text{DM} < m_\text{med} / 2$, the mediator decays dominantly invisibly and di-jet searches are much less sensitive than mono-$X$ searches, whereas in the off-shell region the mediator can only decay visibly and hence di-jet searches give the best constraints.

It is worth noting that searches for di-jet resonances lose sensitivity for low-mass resonances (because of the overwhelming QCD background) and for large widths (because of the uncertainty in the background shape). The former problem can be addressed by searching for di-jet resonances in association with electroweak gauge bosons, which suppresses gluon-induced backgrounds~\cite{An:2012ue,Chala:2015ama,ATLAS:2016bvn}. The latter problem can be addressed by studying not only the invariant mass of the two jets but also their angular distribution~\cite{Khachatryan:2014cja,deVries:2014apa}.

Of course, the mediator may also have further decay modes that can be detectable. For example, a spin-0 $s$-channel mediator with $m_\text{med} > 2 m_t$ would be expected to decay dominantly into top quarks~\cite{Abdallah:2014hon}. Searches for $t\bar{t}$ resonances are however complicated by the fact that both the predicted new-physics signal and the SM background arise from gluon initial states, so that interference effects become important~\cite{Dicus:1994bm,Frederix:2007gi,Djouadi:2015jea,Craig:2015jba}. Because of the resulting peak-dip structure, searches for $t\bar{t}$ resonances are typically less sensitive than searches for di-jet resonances.

Finally, we note that many UV completions of simplified models with spin-1 mediators predict that the mediator couples not only to quarks, but also to charged leptons~\cite{Alves:2013tqa,Arcadi:2013qia,Alves:2015pea,Alves:2015mua,Kahlhoefer:2015bea}. Correspondingly, one could hope to detect the mediator of the DM interactions also in searches for di-lepton resonances~\cite{ATLAS:2016cyf,Khachatryan:2016zqb,CMS:2016abv}. Indeed, these searches are so sensitive, that either the production cross section of the resonance or its branching fraction into leptons must be tiny. In the context of LHC searches for DM it is therefore typically assumed that the mediator does not couple to leptons at high scales and couplings are only introduced at the loop level (if at all)~\cite{Duerr:2016tmh}.

\section{Conclusions}

This review has demonstrated that searching for DM at the LHC is a thriving research field -- both in terms of experimental strategies and theoretical developments. A large number of different mono-$X$ searches are now performed by the experimental collaborations and a number of different theoretical approaches are available for predicting DM signals and interpreting results. A particularly exciting aspect is the complementarity between LHC searches and alternative approaches to the search for DM, which promise to cover the full parameter space of many well-motivated models of DM.

It is clear that a discovery of DM at the LHC is not guaranteed -- many well-motivated DM models like axions or sterile neutrinos predict production cross sections that are much too small to be observable. Nevertheless, the LHC will be able to comprehensively test the paradigm that DM was produced via thermal freeze-out, provided DM couples to quarks or gluons. In other words, if DM ever was in thermal equilibrium with SM states, the potential to discover DM at the LHC is large. Conversely, a non-observation of DM at the LHC would have important implications for our understanding of DM, in particular when combined with the information from other types of DM searches.

While there have been many studies on the potential sensitivity of different search strategies and many proposals for how to compare experimental bounds from different experiments, up to now there is a surprising lack of studies for how to interpret the observation of an excess at the LHC. One explanation for this may be that such an interpretation is inherently difficult: The missing energy spectrum predicted in most DM models is essentially featureless and depends only mildly on the properties of DM. Reconstructing the mass, spin and couplings of the DM particle will therefore likely require input from several mono-$X$ searches, and possibly even the combination of LHC and non-LHC searches for DM.

Simplified models may prove to be a useful tool for such a comparison, but it is also conceivable that further extensions of this framework will be necessary. In any case, it is certainly timely to explore these issues further. The next few years promise spectacular advances in the sensitivity for DM, and we may move very quickly from setting bounds to interpreting hints and finally to mapping out the DM parameter space. No experiment or search channel will achieve this goal in isolation~-- so it is imperative to maintain a broad experimental program and an open mind about the models used to describe the interactions of DM.

\section*{Acknowledgments}

F.~Kahlhoefer would like to thank Sebastian Bruggisser, Alexander Grohsjean, Ulrich Haisch, Enrico Morgante and Kai Schmidt-Hoberg for useful comments on the manuscript.

\end{document}